\documentclass[aps, prb, amsmath, amssymb, english, twocolumn,reprint,floatfix, superscriptaddress]{revtex4-2}
\usepackage[utf8]{inputenc}
\usepackage{graphicx, color}
\usepackage{dcolumn}
\usepackage{bm}
\usepackage[version=4]{mhchem} 
\usepackage{booktabs}
\usepackage{float}
\usepackage{algorithm}
\usepackage{algorithmic}
\usepackage{xcolor}
\usepackage{braket}
\usepackage[colorlinks,urlcolor=blue,linkcolor=blue,anchorcolor=blue,citecolor=blue]{hyperref} 
\preprint{APS/123-QED}

\begin{document}

\title{\textbf{Ultrafast magnetic moment transfer and bandgap renormalization in monolayer \ce{FeCl2}}} 

\author{Yu-Hui Song}\affiliation{School of Physics and Beijing Key Laboratory of Opto-electronic Functional Materials $\&$ Micro-nano Devices, Renmin University of China, Beijing 100872, China}\affiliation{Key Laboratory of Quantum State Construction and Manipulation (Ministry of Education), Renmin University of China, Beijing 100872, China}

\author{Huan-Cheng Yang}\email{hcyang@ruc.edu.cn}\affiliation{School of Physics and Beijing Key Laboratory of Opto-electronic Functional Materials $\&$ Micro-nano Devices, Renmin University of China, Beijing 100872, China}\affiliation{Key Laboratory of Quantum State Construction and Manipulation (Ministry of Education), Renmin University of China, Beijing 100872, China} 

\author{Kai Liu}\email{kliu@ruc.edu.cn}\affiliation{School of Physics and Beijing Key Laboratory of Opto-electronic Functional Materials $\&$ Micro-nano Devices, Renmin University of China, Beijing 100872, China}\affiliation{Key Laboratory of Quantum State Construction and Manipulation (Ministry of Education), Renmin University of China, Beijing 100872, China} 

\author{Zhong-Yi Lu}\email{zlu@ruc.edu.cn}\affiliation{School of Physics and Beijing Key Laboratory of Opto-electronic Functional Materials $\&$ Micro-nano Devices, Renmin University of China, Beijing 100872, China}\affiliation{Key Laboratory of Quantum State Construction and Manipulation (Ministry of Education), Renmin University of China, Beijing 100872, China}\affiliation{Hefei National Laboratory, Hefei 230088, China}      
\date{\today}

\begin{abstract}
The microscopic origin of laser-induced ultrafast demagnetization remains an open question, to which the non-thermal electronic distribution plays a vital role at the initial stage. Herein, we investigate the connection between the non-thermal electronic distribution and the ultrafast spin dynamics as well as the electronic structure evolution in ferromagnetic \ce{FeCl2} monolayer using real-time time-dependent density functional theory (rt-TDDFT) with self-consistent Hubbard $U$ correction. Our simulations reveal that femtosecond laser pulses induce ultrafast magnetic moment transfer from Fe to Cl atoms. More importantly, through a comprehensive analysis of orbital-resolved electronic structure, we elucidate the microscopic origin of this transfer, attributing it to specific intra-atomic and inter-atomic charge transfer pathways driven by non-thermal excitations. The extent of demagnetization of Fe atoms exhibits a non-monotonic dependence on the laser photon energy, reaching a maximum at the resonant excitation. In addition, the dynamical evolution of the band structure was studied based on the eigenstates of the instantaneous Hamiltonian. Under resonant excitation, the bandgap reduction reaches up to $41\%$ within tens of fs. These findings provide fundamental insights into ultrafast spin control and suggest a strategy to optically engineer the magnetism in two-dimensional magnetic materials. 
\end{abstract}

\maketitle 

\section{Introduction}
Since the seminal discovery of ultrafast demagnetization in Ni by Beaurepaire $et$ $al.$ (1996)~\cite{beaurepaire1996ultrafast}, femtosecond laser-driven spin dynamics has emerged as a transformative paradigm, offering novel pathways for high-speed data storage and quantum technologies~\cite{guo2024laser,sato2016laser,huo2025ultrafast}. This phenomenon has been observed in a variety of materials, such as ferromagnetic metals~\cite{ferromagnetic}, ferromagnetic semiconductors~\cite{chen2019revealing} and antiferromagnetic dielectrics~\cite{guo2025ultrafast,nvemec2018antiferromagnetic}. Nevertheless, the microscopic origin of ultrafast demagnetization remains an open question. Several mechanisms have been proposed, including spin-flipping~\cite{gunther2014testing}, spin transport~\cite{malinowski2008control,liu2023microscopic}, non-thermal electronic distribution~\cite{carva2011ab,carva2013ab}, and laser-induced lattice strain~\cite{von2020spin}. From a temporal perspective, ‌the key processes occur across distinct timescales: non-thermal electronic distribution (where laser-excited electrons deviate from Fermi-Dirac statistics) manifests within 100 fs‌; spin-flipping and transport predominantly govern the tens-to-hundreds of femtosecond scale; and ‌lattice strain effects play a major role at longer timescales extending to the picosecond regime. Thus, the laser-induced non-thermal electronic distribution is the key driver of the initial sub-100 fs spin dynamics. Although time-resolved techniques like the magneto-optical Kerr effect (TR-MOKE) and angle-resolved photoemission spectroscopy (TR-ARPES) can probe non-thermal electronic distribution, their roles in quantitatively depicting the microscopic process still remain elusive. This may arise from some limitations in experimental techniques: TR-MOKE lacks sensitivity to nonthermal states~\cite{shim2020role}, while TR-ARPES only probes specific $k$-points~\cite{pierantozzi2024relevance}, lacking the resolution for global electron dynamics. On the theoretical side, phenomenological models often oversimplify energy transfer between subsystems, being unable to capture the full range of microscopic quantum effects~\cite{carpene2006ultrafast}. Consequently, a quantum mechanical description of the laser-induced non-thermal electronic distribution, based on $ab$ $initio$ electronic structure calculations, is essential to achieve a comprehensive understanding of ultrafast demagnetization dynamics. 

A critical question, therefore, is how the non-thermal electronic distribution triggers the initial redistribution of angular momentum. On the ultrashort timescale of laser excitation (within tens of fs), the primary form of angular momentum transfer is an ultrafast flow of spin angular momentum. This typically occurs between localized electrons of magnetic atoms (e.g., $d$-electrons) and delocalized electrons of ligands (e.g., $p$-electrons) or itinerant electrons, driven by the laser-induced non-thermal electronic distribution~\cite{chen2019revealing,Chen2023}. This process subsequently initiates two key processes on longer timescales: the transfer of spin angular momentum to orbital angular momentum via spin-orbit coupling, and the transfer of this orbital angular momentum to the lattice system, thereby impacting the entire demagnetization dynamics~\cite{Tauchert2022,Dewhurst2021,Wu2024}. Hence, a detailed analysis of this initial spin transfer dynamics is crucial for understanding the entire demagnetization cascade in materials.

Two-dimensional van der Waals (2D vdW) magnets, known for their high susceptibility to external stimuli (such as mechanical strain, electric field, magnetic field, and electromagnetic waves‌), provide an ideal platform for exploring the mechanisms of novel spin dynamics~\cite{burch2018magnetism,liu2020light,Zollitsch2023}. Among these 2D vdW magnetic materials, the recently synthesized monolayer \ce{FeCl2} is an air-stable ferromagnetic insulator with a Curie temperature of 147 K and a wide bandgap of 4.2 ± 0.2 eV~\cite{jiang2022general,lu2022unique,zhou2024evidence,aguirre2024ferromagnetic}, which greatly facilitates the experimental investigation. Indeed, the demagnetization dynamics of monolayer \ce{FeCl2} have been investigated experimentally using TR-MOKE measurements over a 400 ps timescale, with a focus on the temperature variation effect~\cite{zhou2024evidence}. On the other hand, the structural simplicity of \ce{FeCl2} makes it also a promising and computationally tractable system to study the spin-charge dynamics and disentangle intricate non-thermal electronic distribution effects. However, the laser-induced spin and charge dynamics on the initial sub-100 fs timescale remain unexplored and require theoretical elucidation. 

In this work, we employ real-time time-dependent density functional theory (rt-TDDFT) combined with the ACBN0 functional to simulate laser-driven spin and charge dynamics in monolayer \ce{FeCl2}. Our simulations specifically probe the role of the non-thermal electronic distribution in ultrafast demagnetization, and reveal ultrafast magnetic moment transfer from Fe to Cl atoms and band structure reconstruction. Furthermore, through a comprehensive analysis of the orbital-resolved electronic structure, we elucidate the microscopic origin of this transfer. Moreover, we investigate the spin
and charge dynamical responses to pumping lasers with different photon energies in order to study the demagnetization processes under distinct experimental conditions.

\section{ Methods and parameters}
TDDFT simulations. The calculations were performed based on time-dependent density functional theory (TDDFT) as implemented in the Octopus package~\cite{oct-2015,tancogne2020octopus}. We employed the local density approximation (LDA) functional with norm-conserving Hartwigsen-Goedecker-Hutter (HGH) pseudopotentials. The key electronic correlation effects were captured using the ACBN0 functional, a pseudohybrid DFT+$U$ method that computes the Hubbard $U$ and Hund's $J$ parameters $ab$ $initio$ and self-consistently~\cite{acbn02015reformulation}. This approach has been extended to real-time TDDFT simulations and yields realistic electronic properties for correlated systems~\cite{tancogne2017self,tancogne2018ultrafast,topp2018all,ilyas2024terahertz,zhang2023ultrafast}. In the time-dependent simulations, the laser field was incorporated via the standard minimal coupling prescription using a time-dependent vector potential $A(t)$. The corresponding electric field is given by $\mathbf{E}(t) = -\frac{1}{c} \frac{\partial \mathbf{A}(t)}{\partial t}$, where $c$ is the speed of light in vacuum. We used a laser pulse with a $\sin^2$ envelope, a full width at half maximum (FWHM) duration of 15 fs, and a carrier-envelope phase of $\phi = 0$. Spin-orbit coupling was incorporated based on the relativistic pseudopotentials.      

The slab model for \ce{FeCl2} took experimental lattices (3.6\,\AA) and atomic positions ~\cite{de1986x} and a 15 Å vacuum region to minimize interactions between periodic images. We adopted a real space grid spacing of 0.153 Å (0.29 Bohr), and 15 × 15 × 1 and 15 × 9 × 1 Monkhorst-Pack k-point meshes for the primitive cell and conventional cell, respectively. 

The time-dependent electron population change. The time-dependent electron population change of a specific band $m$, denoted as $\Delta N_m(t)$, was calculated by projecting the time-evolved Kohn-Sham orbitals $\ket{\psi_{n,k}(t)}$ onto the ground-state Kohn-Sham orbitals $\{\ket{\psi_{m,k}^{\mathrm{GS}}}\}$ and evaluating the absolute deviation from the initial occupation. The expression is given by:
\begin{equation}
\Delta N_m(t) = \frac{1}{N_k} \left| \sum_{k} \left[ \sum_{n \in \text{occ}} \left| \langle \psi_{n,k}(t) \mid \psi_{m,k}^{\text{GS}} \rangle \right|^2  - n_{m,k}^{\text{GS}}\right] \right|,
\label{population}
\end{equation}
where $N_k$ is the number of k-points sampling the Brillouin zone and $n_{m,k}^{\mathrm{GS}}$ is the  ground-state occupation number of band $m$ at $k$. The $\sum_{n \in \mathrm{occ}}$ includes all initially occupied bands.

Orbital- and state-resolved electron transfer. The time-dependent electron transfer from a set of atomic orbitals $\alpha$ (e.g., Fe $3d$, Cl $3p$) of state $n$ to a final state $m$ is given by:
\begin{equation}
Q_{\alpha \to m}^{(n)}(t) = \frac{1}{N_k}\sum_k w_{\alpha,n,k}^\text{GS} \cdot |\langle \psi_{n,k}(t)|\psi_{m,k}^\text{GS}\rangle|^2,
\label{2}
\end{equation}
where $w_{\alpha,n,k}^\text{GS}$ is the projected weight of the atomic orbital $\alpha$ onto the initial state $n$ at $k$ in the ground state.

\section{results and analysis}
\subsection{Ground-state magnetic properties and electronic structure}
\begin{figure}[htbp]
    \centering
    \includegraphics[width=1.0\linewidth]{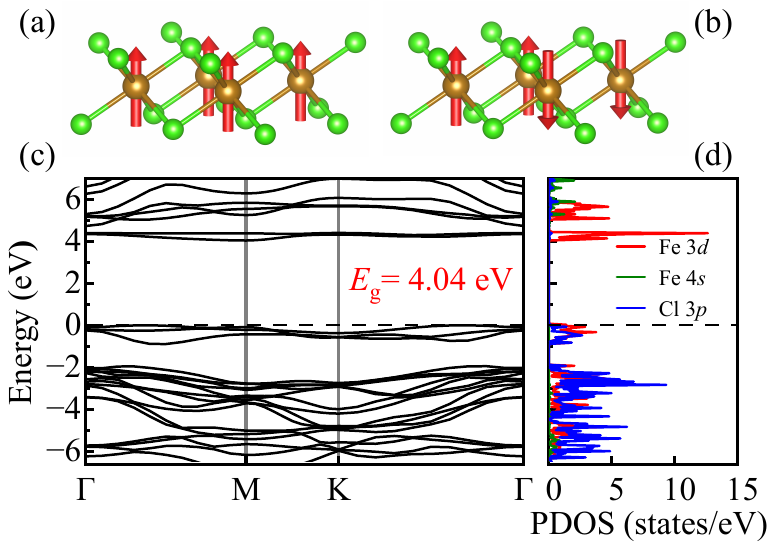}
    \caption{(a, b) Crystal structure and magnetic configurations of \ce{FeCl2}. The arrow's alignment shows the Ferromagnetic (FM) state (a) and antiferromagnetic (AFM) state (b). Golden and green spheres represent Fe and Cl atoms, respectively. (c, d) Electronic structure of the FM configuration: (c) Band structure calculated with Hubbard $U$ correlations and spin-orbit coupling. The valence band maximum (VBM) is set to zero. (d) Partial density of states (PDOS).}
    \label{gs}
\end{figure}

For monolayer \ce{FeCl2}, we consider the correlation effects of Fe 3$d$ and Cl 3$p$ orbitals, and the resulting effective $U_{\text{eff}}$ ($U_{\text{eff}}$ = $U$ - $J$) values are 5.00 eV and 4.45 eV, respectively. The calculated energy of the FM state [Fig. \textcolor{blue}{\ref{gs}}(a)] is about 27 meV/Fe lower than that of the AFM state [Fig. \textcolor{blue}{\ref{gs}}(b)], confirming that \ce{FeCl2} takes the FM ground state, which is in agreement with the experimental results~\cite{zhou2024evidence}. Moreover, the FM state with the Fe magnetic moment along the $z$-axis (out-of-plane direction) has an energy lower by 0.6 or 3.6 meV/Fe compared to those along the $x$-axis or the $y$-axis (in-plane directions), suggesting an out-of-plane easy magnetization axis. The local magnetic moment on the Fe atom is 3.70 $\mu_{\mathrm{B}}$, which indicates that the Fe atom in the octahedral field of Cl atoms is in a high-spin state.

The calculated electronic band structure of \ce{FeCl2} reveals an indirect band gap of 4.04 eV [Fig. \textcolor{blue}{\ref{gs}}(c)], in excellent agreement with the experimental values (4.2 ± 0.2 eV)~\cite{aguirre2024ferromagnetic}. As shown in the partial density of states [PDOS, Fig. \textcolor{blue}{\ref{gs}}(d)] and projected band structures (Fig. \textcolor{blue}{\ref{pband}} in Appendix B), the valence bands exhibit significant contributions from both Fe 3$d$ and Cl 3$p$ orbitals, while the lower-energy conduction bands are dominated primarily by Fe 3$d$ orbitals. Such an electronic structure facilitates the ultrafast magnetic moment transfer from Fe to Cl atoms, which will be elaborated in the following section.  

\subsection{Spin dynamics under resonant excitation}

\begin{figure}[htbp]
    \centering
    \includegraphics[width=1.0\linewidth]{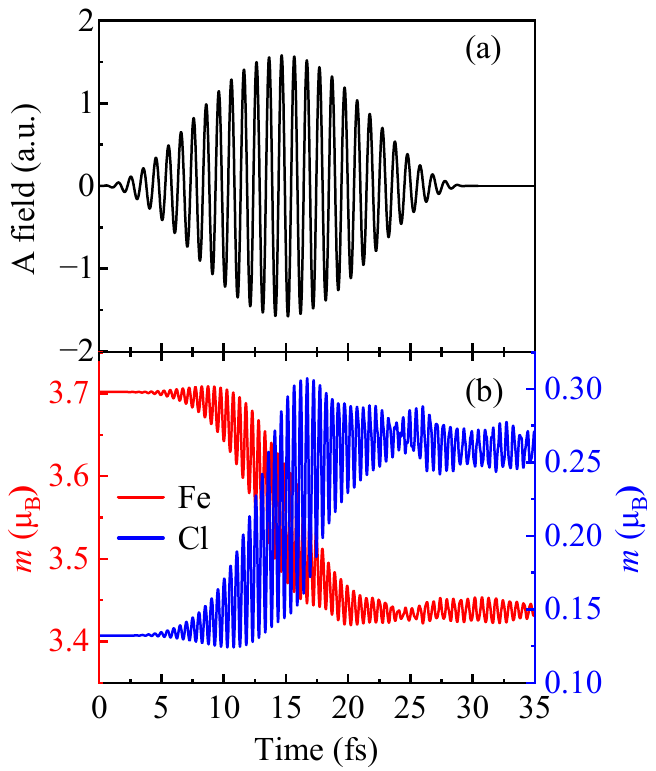}
    \caption{Laser-induced ultrafast spin dynamics in \ce{FeCl2}. (a) Time-dependent vector potential generating the applied in-plane electric fields. The peak amplitude occurs at $t$ = 15 fs. (b) Time evolution of the magnetic moments on the Fe and Cl atoms under resonant excitation at $1.0 \times E_{\text{g}}$.}
    \label{spin}
\end{figure}

In the following, we investigate the photoinduced ultrafast spin dynamics in \ce{FeCl2} under resonant excitation. Fig. \textcolor{blue}{\ref{spin}}(a) displays the temporal profile of the laser electric field for a photon energy equal to the bandgap of \ce{FeCl2} (4.04 eV, i.e., $1.0 \times E_{\text{g}}$). The corresponding pump fluence reaches 12.9 mJ/cm\textsuperscript{2} at a peak intensity of $1.25\!\times\!10^{13}$~W/cm\textsuperscript{2}. Under this resonant excitation, the time evolution of magnetic moments on the Fe and Cl atoms is shown in Fig. \textcolor{blue}{\ref{spin}}(b). For the Fe atom, the magnetic moment decays rapidly during the ultrafast demagnetization phase (0-25 fs) and then oscillates around 3.43 $\mu_{\mathrm{B}}$ in the 25-35 fs time window. The magnetic moment on the Fe atom reduce from 3.70 $\mu_{\mathrm{B}}$ to 3.43 $\mu_{\mathrm{B}}$, showing an effective demagnetization. For the Cl atoms, the time evolution of magnetic moments also exhibits two-stage dynamics similar to those of Fe, while it increases rather than decreases. The final magnetic moments on Cl atoms increase from 0.13 $\mu_{\mathrm{B}}$ to 0.26 $\mu_{\mathrm{B}}$. Collectively, these observations manifest as an ultrafast magnetic moment transfer from Fe atoms to their coordinated Cl ligands.‌

\subsection{Magnetic moment transfer driven by non-thermal electronic distribution}
\begin{figure*}[htbp]
    \centering
    \includegraphics[width=1.0\linewidth]{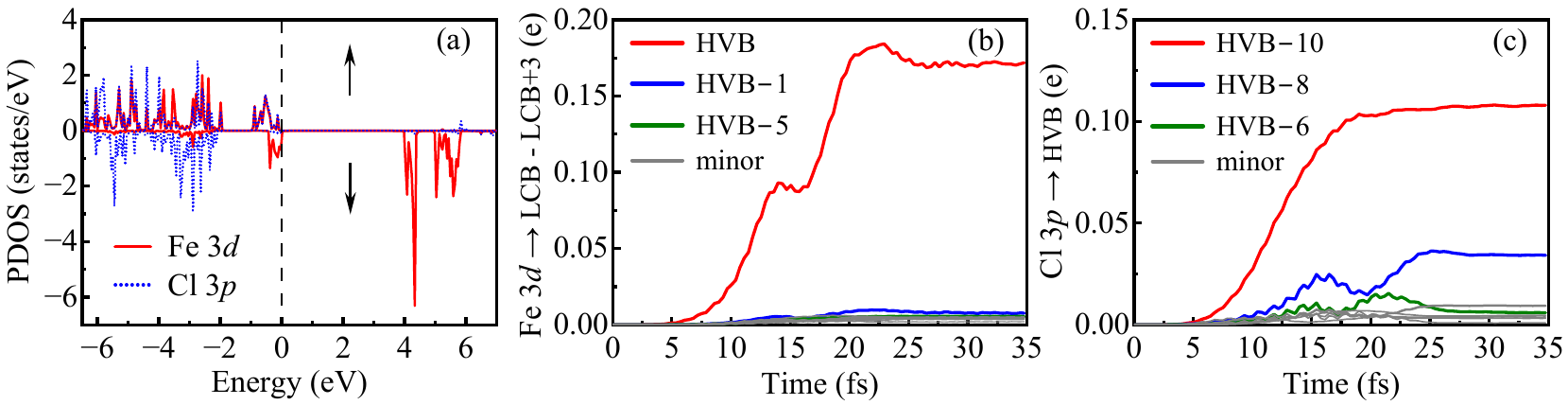}
    \caption{Ultrafast charge transfer pathways in \ce{FeCl2}. (a) Spin-resolved partial density of states at the ground state. The upward and downward arrows represent the spin-up and spin-down channels, respectively. (b) State-resolved electron transfer from Fe $3d$ orbitals to the four lowest unoccupied bands (LCB-LCB+3). (c) State-resolved electron transfer (hole depletion) from Cl $3p$ orbitals to the HVB. The red, blue, and green curves represent the top three contributing bands, while the gray curves represent bands with minor contributions.}
    \label{transfer}
\end{figure*}

To uncover the microscopic mechanism of laser-induced ultrafast magnetic moment transfer from Fe to Cl atoms, we analyzed the charge transfer dynamics under resonant excitation ($1.0 \times E_{\text{g}}$). As shown in Fig. \textcolor{blue}{\ref{transfer}}(a) and Fig. \textcolor{blue}{\ref{pband}} in Appendix B, the spin-projected density of states exhibits pronounced orbital selectivity. The states from HVB to LCB+3 are dominated by Fe 3$d$ spin-down channels, whereas the states from HVB$-$3 to HVB$-$1 have comparable contributions from Fe 3$d$ and Cl $3p$ spin-up channels. Furthermore, the spin-down occupied states in the lower-energy valence bands are mainly contributed by Cl 3$p$ orbitals. (Here, HVB denotes the highest-index valence band, and LCB denotes the lowest-index conduction band. HVB$-$1 refers to the second-highest-index valence band, LCB+1 refers to the second-lowest-index conduction band, and so forth). This electronic configuration provides the foundation for laser-driven inter-orbital spin transfer. The incident laser excites electrons, leading to a non-thermal distribution of electrons within a short time. As shown in Fig. \textcolor{blue}{\ref{transfer}}(b), the HVB band (about $90\%$ Fe 3$d$ spin-down character) transfers approximately 0.18 $e$ to the LCB-LCB+3 bands within 25 fs, which significantly exceeds the contribution from Cl 3$p$ orbitals [$\sim$0.05 $e$, Fig. \textcolor{blue}{\ref{sitransfer}}(a) in Appendix D]. This process drives the intra-atomic magnetic moments (spin angular momentum) reorganization among the $3d$ orbitals of the Fe atom. Conversely, the electron transfer from Cl $3p$ orbitals to the HVB (hole depletion of the HVB) is up to 0.14 $e$ [the sum of the red and blue curves in Fig. \textcolor{blue}{\ref{transfer}}(c)], while electron transfer from Fe 3$d$ orbitals to the HVB remains below 0.01 $e$ [Fig. \textcolor{blue}{\ref{sitransfer}}(b) in Appendix D]. Due to the HVB being almost entirely dominated by Fe spin-down states [Fig. \textcolor{blue}{\ref{transfer}}(a)], the electron transfer from Cl $3p$ to the HVB results in a magnetic moment transfer from Fe to Cl atoms. Our comprehensive analysis thus reveals that Fe 3$d$ → LCB-LCB+3 electron transfer induces intra-atomic spin reorganization within Fe, while hole depletion (Cl 3$p$ → HVB electron transfer) of HVB mediates the magnetic moment transfer from Fe to Cl atoms. These coupled processes mediate the demagnetization of Fe atoms and the concomitant magnetic moment increase of Cl atoms‌, providing direct microscopic support for the above phenomenological statements that “the magnetic moment of Fe atoms rapidly transfers to the coordinated Cl atoms.”
Most importantly, these findings directly implicate laser-excited non-thermal electronic distribution in driving the demagnetization process and allow us to elucidate the specific charge transfer pathways that underlie the observed spin dynamics.

\subsection{Non-monotonic photon-energy dependence of Fe demagnetization}
\begin{figure}[htbp]
    \centering
    \includegraphics[width=1.0\linewidth]{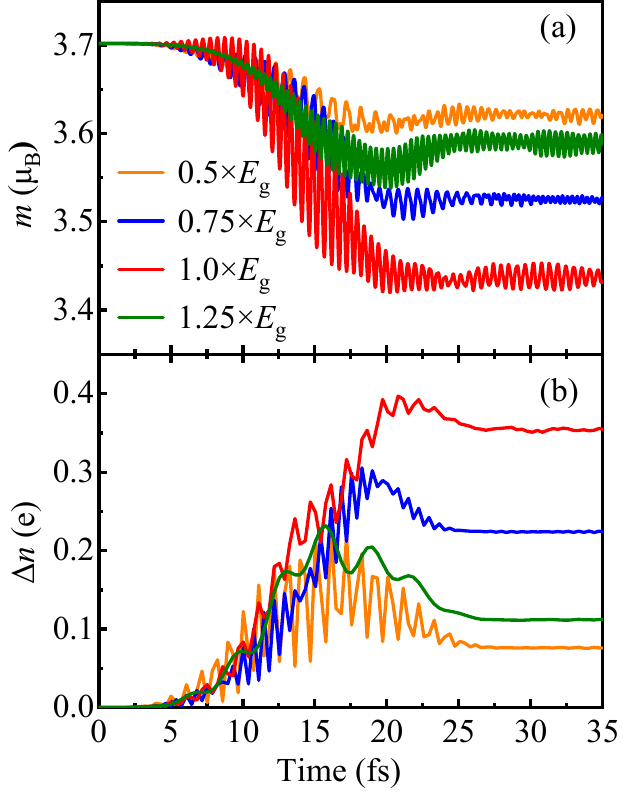}
    \caption{Ultrafast demagnetization of Fe and electron dynamics. (a) Time evolution of the magnetic moments on Fe atoms under laser excitation with four different photon energies: $\hbar\omega=0.5 \times E_{\text{g}}$, $0.75 \times E_{\text{g}}$, $1.0 \times E_{\text{g}}$, and $1.25 \times E_{\text{g}}$, respectively. (b) Total electron population change, $\Delta n = n(t)-n(0)$, summed over the LCB to LCB+3 bands.}
    \label{Fedem}
\end{figure}

\begin{figure*}[htbp]
    \centering
    \includegraphics[width=1.0\linewidth]{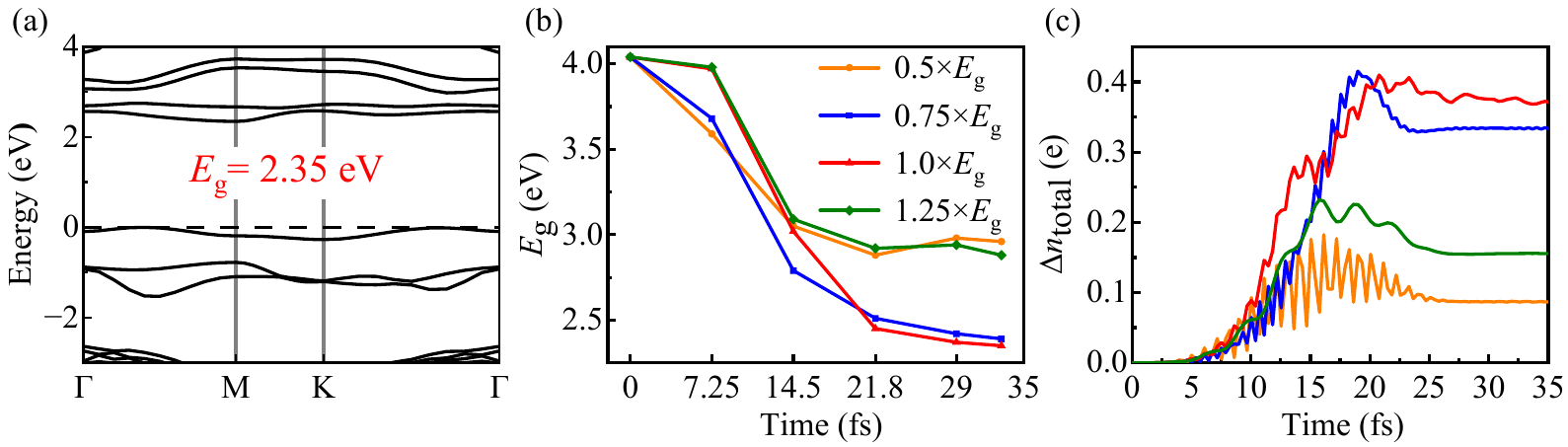}
    \caption{Ultrafast electronic structure dynamics of \ce{FeCl2} under laser excitation. (a) Transient  band structure at $t$=33 fs under resonant excitation at $\hbar\omega=1.0 \times E_{\text{g}}$. (b) Transient bandgap evolution at 0, 7.25, 14.5, 21.8, 29, and 33 fs for different excitation photon energies ($\hbar\omega$=0.5, 0.75, 1.0, and 1.25 $\times E_{\text{g}}$). (c) Time-dependent total electron population change, $\Delta n_{\text{total}}$, summed over the HVB, LCB, and LCB+1 bands, corresponding to the photon energies shown in (b).}
    \label{reband}
\end{figure*}

 Having elucidated the microscopic mechanism of the magnetic moment transfer, we now investigate the photon-energy dependence of Fe demagnetization. As shown in Fig. \textcolor{blue}{\ref{Fedem}}(a), under laser excitation with four different photon energies, the magnetic moments of Fe decay rapidly during the ultrafast demagnetization phase and then oscillate around a specific value. Interestingly, the final magnetic moments at 35 fs show a non-monotonic photon energy dependence: 3.62 $\mu_{\mathrm{B}}$ at $0.5 \times E_{\text{g}}$, 3.52 $\mu_{\mathrm{B}}$ at $0.75 \times E_{\text{g}}$, 3.43 $\mu_{\mathrm{B}}$ at $1.0 \times E_{\text{g}}$, and 3.59 $\mu_{\mathrm{B}}$ at $1.25 \times E_{\text{g}}$. Under resonant excitation, the demagnetization is maximized (a reduction of 0.26 $\mu_{\mathrm{B}}$). To understand this non-monotonic photon energy dependence, we analyzed the electron population [Eq. \eqref{population}] in the LCB-LCB+3 bands. These bands are predominantly composed of Fe 3$d$ spin-down states and thus directly track the key spin-down electron transfer driving demagnetization. The total electron population change summed over the LCB to LCB+3 bands [Fig. \textcolor{blue}{\ref{Fedem}}(b)] reveals significantly enhanced spin-down populations in Fe 3$d$ orbitals, which is a primary driver of the reduction of magnetic moments on Fe atoms (demagnetization). Moreover, the electron population in these bands exhibits the same non-monotonic photon-energy dependence as the Fe demagnetization, confirming that a larger population of Fe 3$d$ spin-down electrons leads to more extensive demagnetization. The matching photon-energy dependence between the Fe demagnetization and the population of spin-down electrons in the LCB-LCB+3 demonstrates that the ultrafast demagnetization is driven by a non-thermal electronic distribution, which leads to the resulting demagnetization of the Fe atom.

\subsection{Laser-induced band reconstruction}
In addition to the ultrafast magnetic moment transfer, the laser excitation also induces ultrafast dynamics in the electronic structure of \ce{FeCl2}. As shown in Fig. \textcolor{blue}{\ref{reband}}(a), the electronic band structure (eigenstates of the instantaneous Hamiltonian) at 33 fs under resonant excitation exhibits three distinct features compared to the initial state [Fig. \textcolor{blue}{\ref{gs}}(c)]. First, the bandgap shows a significant reduction of approximately $41\%$. Second, the LCB+2 and LCB+3 bands undergo a pronounced downshift, which leads to their decoupling from higher-energy conduction bands. Third, the HVB band exhibits a pronounced upshift and decouples from lower-energy valence bands. We also study the dependence of bandgap evolution on photon energy, shown in Fig. \textcolor{blue}{\ref{reband}}(b). Within the first $\sim$ 7 fs, where electronic excitation is negligible, the $E_{\text{g}}$ reduction is closely related to the reduction of $U_{\text{eff}}$. Moreover, for different photon energies during this stage, the reduction of $E_{\text{g}}$ is positively correlated with the decrease in $U_{\text{eff}}$ (Fig. \textcolor{blue}{\ref{Ueff}} in Appendix C). From 7 to about 25 fs, the rate of $E_{\text{g}}$ contraction  becomes co-modulated by both the extent of the $U_{\text{eff}}$ reduction and the growth rate of $\Delta n_{\text{total}}$ [the total electron population change summed over the HVB, LCB, and LCB+1 bands, shown in Fig. \textcolor{blue}{\ref{reband}}(c)]. Subsequently, after $U_{\text{eff}}$ enters an oscillation regime (from 25 to 35 fs), the $E_{\text{g}}$ becomes primarily governed by $\Delta n_{\text{total}}$. At 33 fs, the $E_{\text{g}}$ values are 2.96, 2.39, 2.35, and 2.88 eV for $\hbar\omega=0.5 \times E_{\text{g}}$, $0.75 \times E_{\text{g}}$, $1.0 \times E_{\text{g}}$, and $1.25 \times E_{\text{g}}$, respectively. Therefore, the overall dynamical evolution of the bandgap observed here is a combined result of the laser-induced reduction in electronic correlations and state-filling effects from the non-thermal electronic distribution.

\section{Discussion and Summary}
Based on the above analysis, we confirm the ultrafast spin magnetic moment transfer from Fe atoms to Cl atoms in \ce{FeCl2} under laser excitation. We reveal that this spin magnetic moment transfer originates from the joint processes of Fe 3$d$ → LCB-LCB+3 electron transfer and Cl 3$p$ → HVB electron transfer. Because $p$-orbitals are more spatially extended than $d$-orbitals, the $p$-electrons exhibit a faster spin relaxation rate due to spin-orbit coupling. Furthermore, the spin magnetic moment on $p$-electrons can be transferred more rapidly via mechanisms such as superdiffusive transport~\cite{chen2019revealing,Chen2023}. This efficient dissipation of spin magnetic moment in the $p$-orbitals paves the way for the subsequent rapid decrease of the total spin magnetic moment, i.e., ultrafast demagnetization. Additionally, the laser-induced reduction of the band gap in \ce{FeCl2} facilitates the ultrafast transport of $p$-electron spin magnetic moments.

In summary, using first-principles rt-TDDFT calculations, we study ultrafast spin and charge dynamics in \ce{FeCl2} induced by femtosecond laser pulses. Our study reveals that the demagnetization of Fe atoms and the concomitant magnetic moment increase of Cl atoms are mediated through the synergistic effects of intra-atomic (Fe 3$d$ → LCB-LCB+3) and inter-atomic (Cl 3$p$ → HVB) charge transfer, driven by non-thermal electronic distribution. Furthermore, the extent of demagnetization of Fe atoms exhibits a non-monotonic dependence on the laser photon energy, with a maximum reduction of 0.26 $\mu_{\mathrm{B}}$ under resonant excitation. In addition, we find that the bandgap reaches a maximum reduction of $41\%$ (corresponding to a final value of 2.36 eV) under resonant excitation, and the final $E_{\text{g}}$ for the different photon energies is primarily governed by the total change of electron population ($\Delta n_{\text{total}}$) in the HVB, LCB, and LCB+1 bands. Our work demonstrates a universal approach for optically controlling magnetism in 2D magnetic materials. This strategy not only advances the fundamental understanding of ultrafast spin dynamics but also facilitates the development of photon-driven spintronic devices.

\begin{acknowledgments}
This work was supported by the National Key R\&D Program of China (Grants No. 2024YFA1408601 and No. 2024YFA1408603) and the National Natural Science Foundation of China (Grant No. 12174443). Computational resources
have been provided by the Physical Laboratory of High Performance Computing at Renmin University of China.
\end{acknowledgments}

\appendix

\section{ACBN0 functional implementation}
The time-dependent generalized Kohn-Sham equation within the adiabatic approximation (in atomic units) is given by~\cite{tancogne2018ultrafast}:
\begin{equation}
\begin{split}
i \frac{\partial |\psi_{n,k}^{\sigma}(t)\rangle}{\partial t} &= \Bigg[ -\frac{\nabla^2}{2} + \nu_{\text{ext}}(t) + \nu_{\text{H}}[n(\mathbf{r},t)] \\
& + \nu_{\text{xc}}[n(\mathbf{r},t)] + V_U[n(\mathbf{r},t), \{n_{mm'}^{\sigma}\}] \Bigg] |\psi_{n,k}^{\sigma}(t)\rangle.
\end{split}
\label{ACBN0kS}
\end{equation}
Here, \( \psi_{n,k}^{\sigma}(t) \) denotes a Bloch state with band index \( n \), wavevector $k$ in the Brillouin zone, and spin index \( \sigma \). The potential terms are: \( \nu_{\mathrm{ext}} \) (external potential including laser and ionic potentials), \( \nu_{\mathrm{H}} \) (Hartree potential), \( \nu_{\mathrm{xc}} \) (exchange-correlation potential), and \( V_{U} \) (nonlocal Hubbard correction operator). 
The Hubbard operator  \( V_{U} \) is defined as~\cite{tancogne2018ultrafast}:
\begin{equation}
    V_{U}[n(\mathbf{r},t), \{n_{mm'}^{\sigma}\}] = U_{\mathrm{eff}} \sum_{m,m'} \left( 
\frac{1}{2} \delta_{mm'} - n_{mm'}^{\sigma} \right) P_{m,m'}^{\sigma}, 
\label{ACBN0Vu}
\end{equation}
where \( P_{m,m'}^{\sigma} = | \phi_{m}^{\sigma} \rangle \langle \phi_{m'}^{\sigma} | \) is the projector over the localized subspace spanned by orbitals \( \{ \phi_{m}^{\sigma} \} \), and \( n^{\sigma} \) is the density matrix of this subspace.

\section{Projected band structures}
\begin{figure}[htbp]
    \centering
    \includegraphics[width=1.0\linewidth]{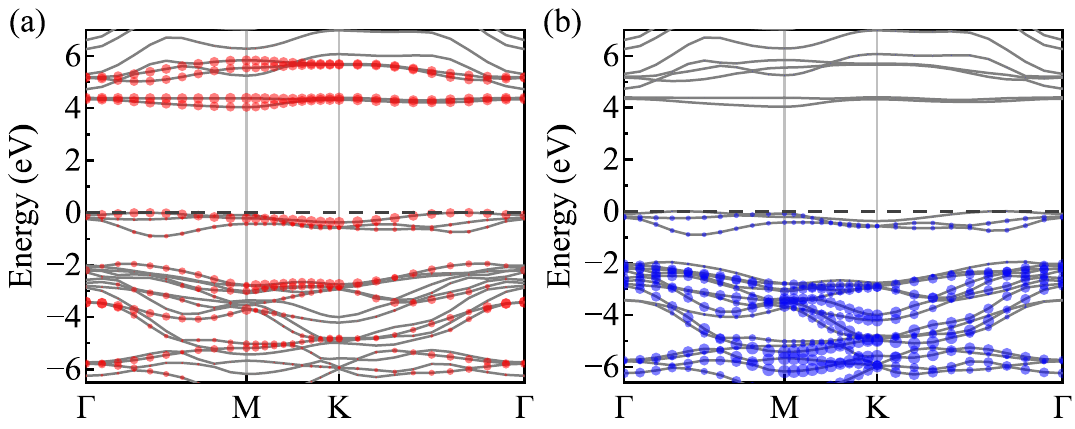}
    \caption{Projected band structures of FeCl$_2$ calculated using the ACBN0 functional. The contributions of (a) Fe $3d$ and (b) Cl $3p$ orbitals are denoted by red and blue dots, respectively. The VBM is set to zero.}
    \label{pband}
\end{figure}

The projected band structures of FeCl$_2$ (Fig. \textcolor{blue}{\ref{pband}}) reveal that the valence bands arise from hybridization between Fe $3d$ and Cl $3p$ orbitals. The HVB and the lowest four conduction bands are predominantly contributed by Fe $3d$ orbitals. 

\section{Dynamics of the effective Hubbard $U$}
\begin{figure}[htbp]
    \centering
    \includegraphics[width=1.0\linewidth]{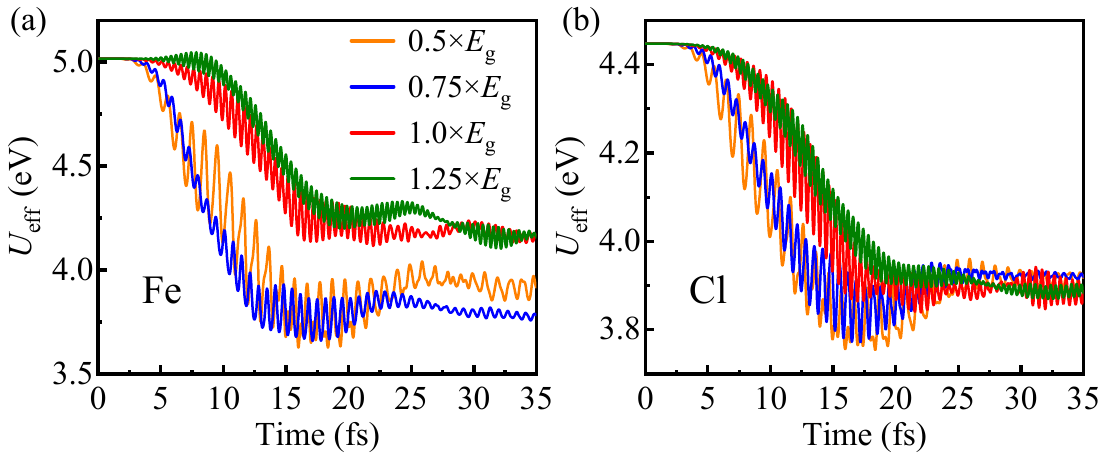}
    \caption{Laser driven dynamics of the effective Hubbard $U$ ($U_{\text{eff}}$). Time evolution of $U_{\text{eff}}$ for (a)  Fe $3d$ and  (b) Cl $3p$  orbitals under $\hbar\omega=0.5 \times E_{\text{g}}$, $0.75 \times E_{\text{g}}$, $1.0 \times E_{\text{g}}$, and $1.25 \times E_{\text{g}}$, respectively.}
    \label{Ueff}
\end{figure}

During ultrafast laser-induced demagnetization, the $U_{\text{eff}}$ evolves dynamically as shown in Fig. \textcolor{blue}{\ref{Ueff}}. For Fe $3d$ orbitals, $U_{\text{eff}}$ decreases to 4.1 eV under $\hbar\omega=1.0 \times E_{\text{g}}$ excitation, with the largest decrease to 3.8 eV occurring at $\hbar\omega=0.75 \times E_{\text{g}}$. For Cl $3p$ orbitals, $U_{\text{eff}}$ decreases from 4.45 eV to approximately 3.9 eV for all four excitation energies, with fluctuations within ±0.15 eV.

\section{Ultrafast charge transfer}
\begin{figure}[b]
    \centering
\includegraphics[width=0.9\linewidth]{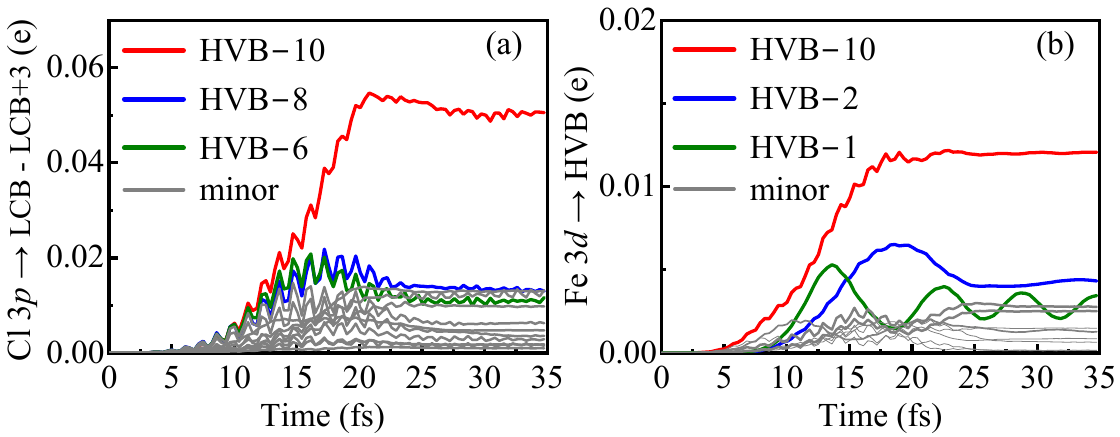}
    \caption{(a) The state-resolved electron transfer from Cl $3p$ orbitals to the LCB-LCB+3. (b) State-resolved electron transfer from Fe $3d$ orbitals to the HVB.}
    \label{sitransfer}
\end{figure}


\bibliography{main}

\end{document}